\documentclass{ceab}   

\usepackage{epsfig}     
\usepackage{graphicx}   

\usepackage{ceabbib}     
\usepackage[T1]{fontenc}

\begin{document}

\def\tit{SW Sex stars and CV evolution}
\def\aut{Linda Schmidtobreick}

\title{The SW Sex phenomenon as an evolutionary stage of Cataclysmic Variables}

\author{Linda Schmidtobreick 
\vspace{2mm}\\
\it European Southern Observatory, Alonso de Cordova 3107,
Vitacura, Santiago, Chile\\
}

\maketitle

\begin{abstract}
From recent large observing campaigns, one finds that nearly all
non- or weakly magnetic cataclysmic variables in the 
orbital period range between 2.8 and 4 hours are of SW Sex 
type and as such experience very high mass transfer rates. 
The evolution of cataclysmic variables as for any interacting binary is driven 
by angular momentum loss which results in a decrease of the 
orbital period on evolutionary time scales. In particular, 
all long-period systems need to cross the SW Sex regime of the
orbital period distribution before entering the period gap. This 
makes the SW Sex phenomenon an evolutionary stage in the 
life of a cataclysmic variable. Here, I present a short overview of 
the current state of research on these systems.
\end{abstract}


\section{The evolution of cataclysmic variables}
Cataclysmic variables (CVs) are semi-detached binaries comprising a Roche
lobe filling low-mass main-sequence star transferring
matter onto a white dwarf.
The evolution of any CV is determined by the
loss of angular momentum it experiences. 
Without any braking mechanism, the angular momentum is conserved in the binary.
Hence, the system will spin up when mass 
is transferred towards the more massive 
primary and thus closer to the centre of mass. This will eventually 
result in the secondary getting detached from its Roche-lobe.
To establish a stable mass-transfer, the continuous loss of angular 
momentum is fundamental. The CVs are thus evolving from longer orbital periods
towards shorter orbital periods. The standard model predicts 
magnetic braking as the dominant source for angular momentum loss 
for long-period CVs ($P\ge 3$\,h) while CVs below the period gap 
($P\le 2$\,h) are supposed to lose angular momentum via 
gravitational radiation only. The special case of CVs that reached the
orbital period minimum and in which the secondary becomes degenerate has
no impact on the SW\,Sex phenomenon and is thus no further discussed.  

\citet{rappaportetal83-1} suggested
the scenario of disruptive magnetic braking with the assumption that
the magnetic braking is efficient as long as the secondary star has a 
radiative core but becomes negligible when the secondary star is fully 
convective at about $M_2 \approx 0.25 M_\odot$. For a Roche-lobe filling
secondary star, this mass corresponds to an orbital period of about 
$P =3$\,h, i.e. just at the upper edge of the period gap. The secondary
star is driven out of thermal equilibrium due to the large mass loss rate
and is therefore oversized for its mass. Once the magnetic braking
stops, the mass loss rate drops and the star will relax to its 
normal size in thermal equilibrium, thus losing the contact with its
Roche-lobe. The binary is now detached and not detectable as a CV.
It is still evolving towards shorter orbital periods by losing angular 
momentum via gravitational radiation. At an orbital period of about 2\,h
the Roche-lobe is sufficiently small that mass transfer will be established
again but on a low level driven by gravitational braking only.

Even though it is not yet understood why the magnetic braking would cease
when the secondary becomes fully convective, the standard model is 
widely accepted, not least because there exist several pieces of
observational evidence for 
the disruptive magnetic braking scenario. 
\citet{pattersonetal05-3} showed from superhump measurements that the 
donor mass of CVs at the upper edge of the gap seems to be similar to
the donor mass of CVs just below the gap. This supports the idea that 
the period gap is indeed a region in which no mass transfer happens and 
through which the CVs evolve as detached systems. The second evidence comes
from the observations that the 
donors above the gap are more bloated than donors below the gap indicating
that they experience more mass loss which drives them out of thermal 
equilibrium \citep{kniggeetal11-1}. Third, using the temperature of 
white dwarfs as a measure for the
accretion induced compressional heating \citet{Town+09} could show that 
the accretion rate above the gap is higher than the one below the gap, and
last but not least, \citet{schreiberetal2011-1} showed that the mass
distribution of post common envelope binaries is in agreement with the
predictions of \citet{politano+weiler06} for disrupted magnetic braking.

Observationally, there is thus no doubt for a discontinuity in the mass
transfer rate of CVs around an orbital period of 3\,h but the reason for
the reduced magnetic braking is still under discussion. 

\section{SW Sextantis stars}
SW Sextantis stars are a group of CVs 
that were originally defined as eclipsing nova-like stars which
show single-peak emission lines, high-velocity emission line wings, 
strong He\,II emission but no polarisation, and transient absorption 
features in the emission lines at an orbital phase of $\phi = 0.5$
\citep{thorstensenetal91-1}. They also show orbital phase offsets 
of 0.2 cycles of the radial velocity curves with respect to the 
photometric ephemeris and are interpreted as having high
mass transfer rates (see e.g. \citet{rodriguez-giletal07-1}). 
The idea that these binaries have high mass transfer rates
is supported
by the study of \citet{Town+09} who show that the white dwarf temperature
of these stars exceeds the expected value for accretion governed by an angular 
momentum loss from standard magnetic braking.

Initially, SW Sex stars were considered as odd objects. However,
\citet{gaensicke05-1} showed that they are instead common in the orbital 
period range between 2.8 and 4\,h, i.e. the range just above the period gap.
In particular, all the eclipsing nova-like systems in this period range
belong to the sub-class of SW Sex stars.

Being an eclipsing system is, however, not an
intrinsic physical property of the star but rather
depends on the angle under which the binary is
observed. It thus appears entirely plausible that all
non- or weakly-magnetic CVs just above the
period gap are physically SW Sex stars, i.e. experience a very 
high mass transfer rate. 

To test this hypothesis, \citet{2007MNRAS.374.1359R} and
Schmidtobreick et al. (in preparation) have conducted a survey and 
obtained time-series spectroscopy on the non-eclipsing CVs with 
$V\le 18$ in the 2.8--4\,h period range. 
They searched for the presence of the defining SW Sex characteristics
such as broad line wings with large-amplitude radial
velocity variations, single-peaked line profiles
with phase-dependent central absorption, and
phase lags between the radial velocity modulation 
in the line cores and wings. They also checked
for line flaring, an additional feature that is often 
observed in SW\,Sex stars but also in intermediate polars 
and that is manifested in fast oscillations of the 
emission line flux and velocity
with periods around 10-20 min. 

As the main result, they found that indeed the majority of the observed
CVs in the 2.8--4\,h period range, and all nova-like stars among these
are of SW\,Sex type. Several of these are showing the
line-flaring phenomenon. The detailed results of
the campaign are discussed in \citet{2007MNRAS.374.1359R} and
Schmidtobreick et al. (in preparation). They conclude that SW\,Sex stars are 
not oddballs but present instead the major population of CVs in the 
period range just above the gap.

\section{SW\,Sex stars as an evolutionary state}
\begin{figure}
\begin{center}
\includegraphics[scale=1.7]{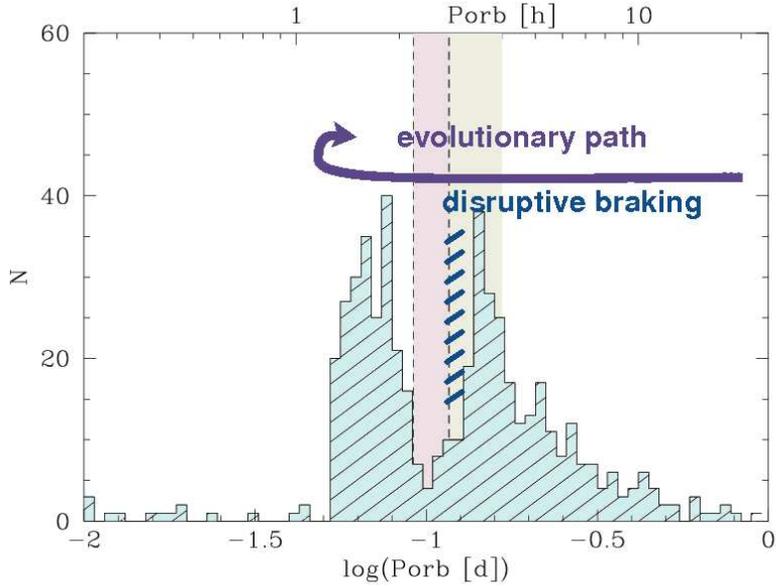}
\end{center}
 \caption{\label{gap} The orbital period distribution as presented by \citet{gaensicke05-1}.
The regime of the SW\,Sex stars and the 
orbital period gap are marked in different colour shades. 
The evolutionary path of a CV due
to angular momentum loss and the position where the magnetic braking
gets reduced are indicated.}
\end{figure}

In the previous sections, I have discussed two statements: (1) CVs
 evolve from longer orbital periods to shorter ones due to 
angular momentum loss, and (2) most CVs and i.e. all nova-likes 
in the period range
from 2.8--4\,h are of SW\,Sex type. Combining these two facts one can deduce
that those CVs that get into contact at longer orbital periods
$P\ge 4$\,h have to evolve through the SW\,Sex regime before they enter
the period gap and most likely become an SW\,Sex star during this phase 
(see Fig.\ \ref{gap}). 
This implies
that the SW\,Sex phenomenon is actually a state in the evolution of CVs. 
More importantly, it is the state of a CV 
just before it enters the period gap, i.e. the state in which a CV is 
when the magnetic braking stops and the binary 
becomes detached.

To understand the evolution of CVs and in particular the evolution into
the orbital period gap, one needs to understand the SW\,Sex phenomenon. 
Models that predict or simulate the evolution of CVs and that make use 
of the magnetic braking and its disruption, also have to 
take into account that before the binary becomes detached, the CV is most
likely an SW\,Sex star and as such experiences a high accretion rate 
that cannot be explained by standard magnetic braking. 

A special case in this context represent the two old novae XX Tau and
V728 Sco, with orbital periods of 3.3 h and 3.32 h, respectively 
\citep{rodriguez-gil+torres05-1, tappertetal12-2}.
Most old novae are high mass-transfer systems \citep[e.g.][]{ibenetal92-2},
and since their orbital periods place these two 
objects right within the SW Sex regime, they have double reason to fall
into this category as does, e.g., the old nova RR Pic 
\citep{schmidtobreicketal03-1}.
Instead, spectroscopic observations indicate that XX Tau and V728 Sco 
have significantly lower mass-transfer rates and share the spectroscopic 
characteristics of dwarf novae rather than nova-likes \citep{schmidtobreicketal05-1, tappertetal12-1}. In fact,
at least V728 Sco has been observed to undergo dwarf-nova like outbursts
\citep{tappertetal12-2}.
This behaviour might
be connected to the nova outburst that these CVs experienced in the past.
The presence of low mass-transfer 
novae in the SW Sex regime could potentially be explained within the context
of the hibernation model \citep{sharaetal86-1, prialnik+shara86-1}, but an
in-depth analysis of the behaviour of old novae with orbital periods
between 2.8\,h and 4\,h is still pending.

So far, most models do not include the SW\,Sex phenomenon, simply because
SW\,Sex stars have so far been considered as weird objects among the 
normal CVs. 
There was just no need to complicate an evolutionary
model only to explain a few outliers. 
However, \citet{zangrillietal-1} have developed 
a model for the magnetic braking that could indeed explain the high mass
transfer rate for systems just above the period gap. They use two 
$\alpha - \Omega$ dynamos, one in the convective envelope and one at 
the boundary layer with a slowly rotating radiative core. The main difference 
to other models is the slow rotation of the core which is considered to
not be synchronised with the orbital motion. During the evolution of the CV,
the core becomes smaller, and its material has to speed up to move with the
synchronised convective envelope. This yields an additional braking which 
becomes stronger towards the fully convective boundary and would explain
an increased mass transfer rate for CVs just above the period gap, 
i.e. the SW Sex stars.

One problem in the understanding of SW\,Sex stars as a group lies
in the fact that only for a few of them, the system parameters like
masses or temperature or even mass ratio is know. This is due to their
high mass transfer rate and the dominating accretion disc that does not allow
a direct observation of either of the binary components. 
Fortunately, most if not all SW Sex stars sometimes switch into a 
low state where their brightness drops by several magnitudes. 
These so-called VY\,Scl low states are rare and unpredictable 
but when they occur
the  mass transfer is strongly reduced or even completely suppressed.
Therefore, the disc becomes much fainter or dissolves completely, and the
stellar components of the binary become accessible 
\citep[see][for a review]{rodriguez-giletal11}.
During such low states, some information on the system parameters
were gathered on TT\,Ari 
\citep{gaensickeetal99-1}, MV\,Lyr \citep{hoardetal04-1}, and
DW\,UMa \citep{kniggeetal00-1,araujo-betancoretal03-1}.
They all show indication for high temperature white dwarfs
consistent with the high mass transfer rates that are expected for these binaries. To increase 
the knowledge on these nova-likes, a project has been started by 
\cite{rodriguez-giletal11} to monitor the nova-like stars photometrically 
and to trigger time-resolved spectroscopic observations as soon as they go into a deep low state. First
results have been obtained for BB\,Dor \citep{rodriguez-giletal12-1, schmidtobreicketal12-1}, VY\,Scl (Schmidtobreick et al.\ in preparation), and 
HS\,0220+0603 (Rodr\'\i guez-Gil et al. in preparation).

\section*{Summary}
I have discussed the two statements that due to angular momentum loss,
CVs evolve from 
longer orbital periods to shorter orbital periods and that almost
all CVs with an orbital period between 2.8\,h and 4\,h are of SW\,Sex type.
Putting both information together, one can conclude that CVs that are
born with an orbital period $P\ge 4$\,h have to evolve through the 
2.8--4\,h period range before entering the period gap. During this phase,
they will almost certainly exhibit the SW\,Sex behaviour. This makes
the SW\,Sex phenomenon an evolutionary state of CVs and in particular 
the evolutionary state during which the magnetic braking gets disrupted. 
SW\,Sex stars thus play an important role in our understanding of 
CV evolution and should be part of any model that tries to
explain or to simulate CV evolution. 
The presence of non-magnetic old novae with low mass-transfer rates in the 
SW Sex regime appears somewhat conflicting in this context. However, it is
well possible that their current state is a consequence of the nova eruption,
as outlined by the hibernation model. 
To include SW\,Sex stars into the
overall evolutionary tracks of CVs, more observational information is needed
for this type of CVs, i.e. the determination of system parameters for
a representative amount of these binaries. However, 
due to the bright accretion disc being the dominating source for emission
in these high mass-transfer systems, information on the stellar components
can only be obtained during one of the rare low states. Consequently,
respective results are still sparse, but since there are a number of projects
dedicated to this research, the situation is likely to improve in the not
too distant future.
\section{Acknowledgements}
I would like to thank Antonio Bianchini, Boris G\"ansicke, Elena Mason, 
Pablo Rodr\'\i guez-Gil, Matthias Schreiber, and Claus Tappert for 
valuable discussions. I also thank the STScI for the
hospitality during the scientific stay (supported by the ESO DGDF program)
in which part of this investigation was done.

\bibliographystyle{ceab}


\end{document}